\documentclass[aps,pre,preprint,showpacs,superscriptaddress]{revtex4}
\usepackage{amsmath}
\usepackage{epsfig}

\preprint{APS/123-QED}

\newfont{\logo}{logo10}
\newcommand{\bear}{\begin{eqnarray}}
\newcommand{\eear}{\end{eqnarray}}
\newcommand{\bes}{\begin{subequations}}
\newcommand{\ees}{\end{subequations}}
\newcommand{\al}{\alpha}

\newcommand{\s}{\sigma}

\begin{document}
\bibliographystyle{revtex}
\title{Periodic energy switching of  bright solitons in mixed coupled nonlinear Schr{\"o}dinger equations with linear self and cross coupling terms}  
\author{T. Kanna\footnote{ e-mail: tkans@rediffmail.com}} 
\affiliation{Department of Physics,
 \\
Bishop Heber College,Tiruchirapalli--620 017, India   }
\author{M. Vijayajayanthi\footnote{ e-mail: jayanthi@cnld.bdu.ac.in}}
\affiliation{Centre for Nonlinear Dynamics, School of Physics, Bharathidasan University, Tiruchirapalli-620 024, India\\} 
\author{M. Lakshmanan \footnote{ Corresponding author e-mail: lakshman@cnld.bdu.ac.in}}
\affiliation{Centre for Nonlinear Dynamics, School of Physics, Bharathidasan University, Tiruchirapalli-620 024, India\\} 
\begin{abstract}

The bright soliton solutions of the mixed 2-coupled nonlinear Schr{\"o}dinger
(CNLS) equations with linear self and cross coupling terms have been obtained by
identifying a transformation that transforms the corresponding equation to the
integrable mixed 2-CNLS equations. The study on the collision dynamics of bright
solitons shows that there exists periodic energy switching, due to the coupling
terms.  This periodic energy switching can be controlled by the new type of
shape changing collisions of bright solitons arising in mixed 2-CNLS system,
characterized by intensity redistribution, amplitude dependent  phase shift and
relative separation distance.  We also point out that this system exhibits large
periodic intensity switching even with very small linear self coupling
strengths. 

\end{abstract} 

\maketitle
\section{Introduction}
The study on soliton collisions has been receiving sustained
attention since the identification of their particle like
collision behaviour by Zabusky and Kruskal \cite{ref1} in 1965. Due
to their intriguing collision properties and their robustness
against external perturbations solitons find applications in
diverse areas of science which encompass the current thrust
research areas including nonlinear optics, Bose-Einstein condensates,
plasma physics, bio-physics \cite{ref2,ref3,ref4,ref5}. Particularly, it has been shown that soliton
propagation in systems such as birefringent fibers \cite{ma,ref6}, multimode fibers
\cite{ref2},
fiber couplers \cite{ref2}, photorefractive medium \cite{ref7}, 
left handed materials \cite{ref8}, Bose-Einstein condensates \cite{ref8a} and continuum limit of 
Hubbard model in 1D \cite{ref21}
is governed by multicomponent nonlinear Schr{\"o}dinger type
equations which become integrable for specific choices of system parameters \cite{ref9,ref10,ref11}.

Recent theoretical and experimental studies show that the 
integrable CNLS system  with focusing type nonlinearity exhibits
fascinating shape-changing (energy/intensity redistributing)
collision of bright solitons characterized by intensity redistribution among the
colliding solitons in all the components, as well as amplitude dependent
phase-shifts and change in relative separations distances \cite{ref12,ref13,ref14}.  In such
a  two soliton collision process there occurs suppression/enhancement
of intensity in few components and enhancement/suppression of intensity in the
remaining components after interaction. We call such a collision
scenario as type-I shape changing collisions (type-I SCC). This interesting collision behaviour has also been experimentally verified in birefringent fibers \cite{ref14c} and in photorefractive media \cite{ref14b}. Further
investigations on this type-I SCC has led to exciting applications in collision based optical computing \cite{ref14,ref15,ref16,ref17}.
However the collision scenario is different if the nonlinearities
are of mixed type \cite{ref19,ref20},  which includes both focusing and
defocusing types. The corresponding mixed CNLS equations admit
shape changing collision of bright solitons in a quite
different manner from the collision scenario of type-I SCC. Very
recently, it has been shown that in mixed CNLS equations during a
two soliton collision process  there is a possibility of either
enhancement or suppression of intensity in a given soliton in all
the components \cite{ref20}. Here also the collision process is characterized
by intensity redistribution, amplitude dependent phase-shift and
relative separation distances. We denote this kind of collision
scenario as type-II shape changing collision (type-II SCC).  The most important consequence of type-II SCC is the
possibility of {\it{soliton amplification}} in all the components.

Mixed CNLS equations are not only of mathematical interest
but also of considerable physical significance. In particular, mixed 2-CNLS
equations arise as governing equations for electromagnetic pulse
propagation in left handed materials with Kerr-type nonlinearity \cite{ref8} and in the modified Hubbard model in the long-wavelength approximation \cite{ref21}. It can also be noticed that in the mixed 2-CNLS system, if we assume the fields $q_1$ and $q_2^*$ (see Eq. (1) below) propagate in the anomalous and normal dispersion regimes, respectively, the self phase modulation (SPM) coefficients are positive and cross phase modulation (XPM) coefficients are negative in both the  components. This kind of nonlinearities can be realized in the quadratic nonlinear materials with large phase-matching \cite{ref23}.
The main aim of this
paper is to analyse type-II SCC behaviour of bright solitons in
a possible integrable extension of mixed 2-CNLS equations involving the
linear self and cross coupling terms which can be of physical
interest. This study reveals the fact that in this system during the type-II SCC process there also occurs periodic energy switching due to the linear coupling terms which can be suppressed/enhanced by type-II SCC. The distinct feature is that the system exhibits large periodic energy switching between the components with very small linear self coupling strength.

The plan of the paper is  as follows.  In Sec. II we present the statement of the problem. Bright soliton solutions of the mixed 2-CNLS equations with linear self and cross coupling terms are obtained in Sec. III.  Section IV is devoted to a detailed analysis of the role of the coupling terms on type-II SCC. The final section is allotted for conclusion.

\section{Statement of the problem} 
Lazarides and Tsironis \cite{ref8} have obtained the following set of 
governing equations for electromagnetic pulse propagation in
isotropic and homogeneous nonlinear left handed materials,
\bes
\begin{eqnarray}
iq_{1,z}+q_{1,tt}+2 \mu\left(\sigma_1 |q_1|^2+\sigma_2 |q_2|^2
\right)q_1 = 0, \\ iq_{2,z}+q_{2,tt}+2 \mu\left(\sigma_1
|q_1|^2+\sigma_2 |q_2|^2 \right)q_2 = 0,
\end{eqnarray}
\ees
by taking the effective permittivity and effective permeability to
be intensity dependent and following a reductive perturbational
approach . Here $q_1$ and $q_2$ are the electric and magnetic field components of the electromagnetic pulse, respectively, the
subscripts $z$ and $t$ denote the partial derivatives with respect to
normalized distance and retarded time respectively, while $\mu$ is the measure of
nonlinearity,  $\sigma_1$ and $\sigma_2$ can be either +1 or -1.
From a mathematical point of view the above equation reduces to
the integrable mixed CNLS equation presented in Ref. \cite{ref10} when
$\sigma_1= 1$ and $\sigma_2=-1$. In Ref. \cite{ref20}  singular and nonsingular bright soliton
solutions of Eq. (1)  have been obtained. There it has been shown that even though
system (1) admits SCC as in Manakov system, the intensity
redistribution occurs in a completely different way which is not
possible in the Manakov system \cite{ma}.  As pointed out in the introduction
we denote such a collision picture as type-II SCC.  A typical type-II SCC where 
 enhancement (suppression) of intensity occurs in both the
components of soliton $S_1$($S_2$) after collision is shown in Fig. 1  (All the quantities in this and rest of the figures are dimensionless). Note that the reverse scenario is also possible.
 \begin{figure}
\begin{center}
\epsfig{figure=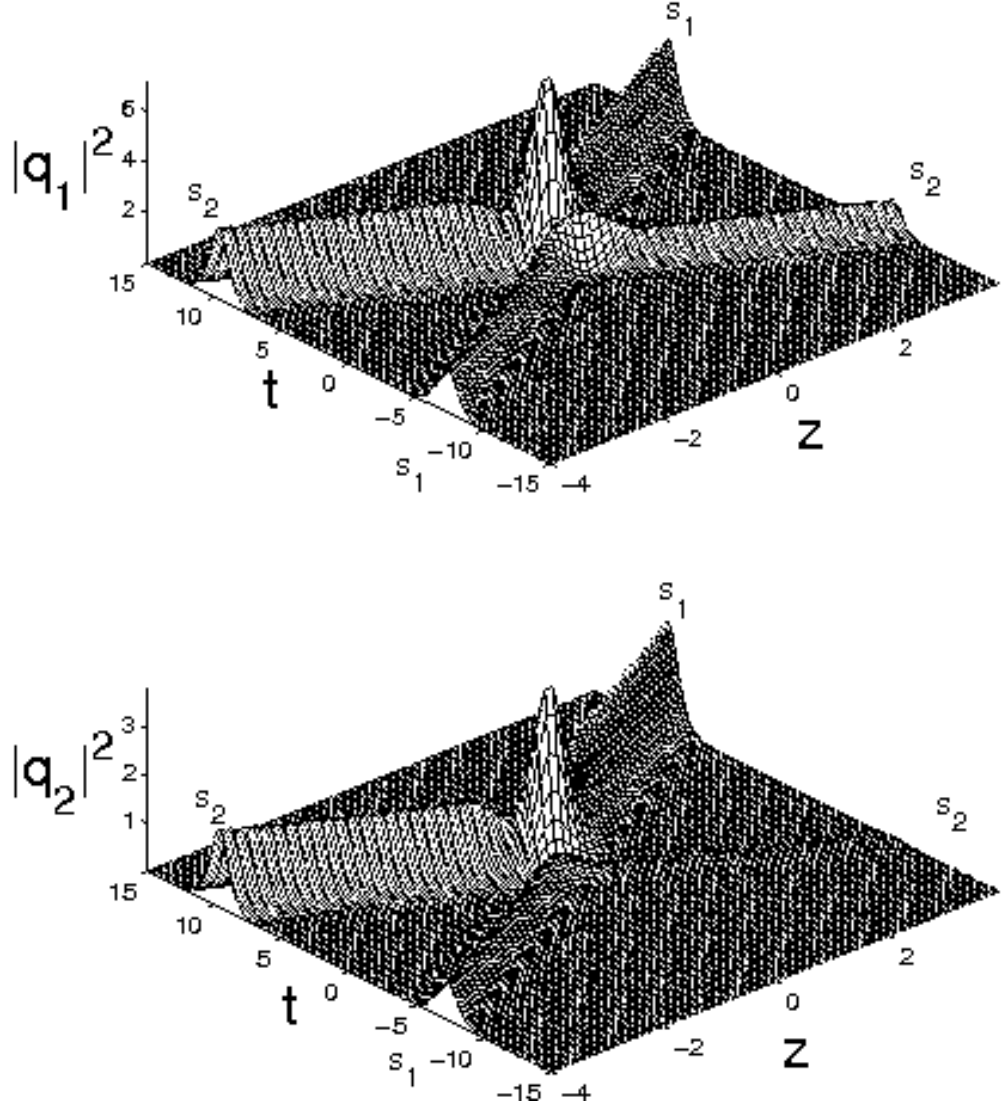, width=0.8 \columnwidth}
\caption{Shape changing collision of bright solitons in mixed 2-CNLS system.}
\end{center}
\end{figure}

The next natural step is to study how the linear self coupling
resulting from the same component and the cross coupling arising
from the other component influence the type-II SCC process. In
this regard we consider the following non-dimensional mixed CNLS equations with
linear self and cross couplings.
\bes
\begin{eqnarray}
iq_{1z}+q_{1tt}+\rho q_1-\chi q_2+2 \mu\left( |q_1|^2- |q_2|^2
\right)q_1 = 0, \\ iq_{2z}+q_{2tt}- \rho q_2 +\chi q_1 +2
\mu\left( |q_1|^2-|q_2|^2 \right)q_2 = 0,
\end{eqnarray}
\ees
where $\rho$ and $\chi$ are the self and cross coupling coefficients,
respectively.  The above system reduces to the Lindner-Fedyanin system \cite{ref21} with $\chi
= 0$, which is a 1D continuum limit of 2D Hubbard model.  We also believe that this mathematical model can be
experimentally realized in fiber couplers fabricated from suitably
engineered left handed materials. 
Another physical model associated with Eq. (2) is two component BECs coupled with two photon microwave field where the signs of co-efficients of SPM and XPM can be tuned suitably through Feshbach resonance. In fact, this type of linear couplings can be introduced in BECs by external microwave or radio frequency that induces Rabi or Josephson oscillation between population of two states \cite{ref24}. Also at this point it is worth
to point out that a similar set of equations with focusing
nonlinearity $(|q_1|^2+ |q_2|^2)$ arises in pulse propagation in
twisted birefringent fibers \cite{ref2, ref22}.  

\section{Bright soliton solutions}
To obtain the bright soliton solutions of Eq. (2), we identify a transformation which reduces Eq. (2) to the well
known integrable mixed CNLS equations (1) with $\sigma_1=1$ and
$\sigma_2=-1$. The transformation can be written as
\bes
\begin{eqnarray}
\left(
\begin{array}{c}
  q_1 \\
  q_2
\end{array}
\right)= \left(
\begin{array}{cc}
  \cosh(\frac{\theta}{2})e^{i\Gamma z} & \sinh(\frac{\theta}{2})e^{-i\Gamma z}\\
  \sinh(\frac{\theta}{2})e^{i\Gamma z} & \cosh(\frac{\theta}{2})e^{-i\Gamma z}
\end{array}
\right)\left(
\begin{array}{c}
  q_{1m} \\
  q_{2m}
\end{array}
\right),
\label{oe}
\end{eqnarray}
where the real parameters $\theta$ and $\Gamma$ are expressed in terms of the coupling
parameters $\rho$ and $\chi$ as
\begin{equation}\theta =
\tanh^{-1}\left(\frac{\chi}{\rho} \right),\;\; \Gamma =
\sqrt{\rho^2-\chi^2}, \;\;\;\chi\leq \rho.
\end{equation} 
\label{reln}
\ees
In Eq. (\ref{oe}), $q_1$ and $q_2$ satisfy Eq. (2) with linear self and cross coupling terms included while $q_{1m}$ and $q_{2m}$ satisfy Eq. (1) in the absence of linear couplings ( $\rho=\chi=0$).  It is
obvious that if the  cross coupling term $\chi$ becomes zero
the above  solution $(q_1, q_2)$ is the same as that of the mixed CNLS
equations (1), with $\sigma_1=-\sigma_2=1$, except for a
multiplicative phase factor $e^{i\Gamma z}$ in the $q_1$ component and
$e^{-i\Gamma z}$ in the $q_2$ component.  We now confine our analysis to the cases where $(q_{1m}, q_{2m})$ correspond to bright soliton solutions and analyse the nature of $(q_1,q_2)$ satisfying Eq. (2) through the relation (\ref{reln}).
\subsection{Bright one-soliton solution}
With the knowledge of the bright one soliton solution of
the integrable mixed 2-CNLS equations for $(q_{1m}, q_{2m})$, given in Eq. (6) of Ref.  \cite{ref20}, we
write down the one soliton solution of Eq. (2) by using the
transformation (\ref{reln}) as
\bes
\begin{eqnarray}
\left(
\begin{array}{c}
  q_1 \\
  q_2
\end{array}
\right)= \left(
\begin{array}{cc}
  \cosh(\frac{\theta}{2})e^{i\Gamma z} & \sinh(\frac{\theta}{2})e^{-i\Gamma z}\\
  \sinh(\frac{\theta}{2})e^{i\Gamma z} & \cosh(\frac{\theta}{2})e^{-i\Gamma z}
\end{array}
\right)\left(
\begin{array}{c}
  A_{1} \\
  A_{2}
\end{array}
\right) k_{1R} \mbox{sech}\left(\eta_{1R}+\frac{R}{2}
\right)e^{i\eta_{1I}},
\end{eqnarray}

where 
\begin{eqnarray}
\eta_1&=&k_1(t+ik_1z)=k_{1R}(t-2k_{1I}z)+i (k_{1I}t+(k_{1R}^2-k_{1I}^2)z) \equiv \eta_{1R}+i\eta_{1I},\\
 A_j&=&\frac{\alpha_1^{(j)}}
{\left[\mu\left(\s_1|\al_1^{(1)}|^2+\s_2|\al_1^{(2)}|^2\right)\right]^{1/2}}, \;\;j=1,2, \\
e^R &=&\frac{\mu\left(\s_1|\al_1^{(1)}|^2+\s_2|\al_1^{(2)}|^2\right)}
{(k_1+k_1^*)^2},\;\; \s_1=-\s_2=1.
\end{eqnarray}
\ees
In the above expressions the suffices R and I denote the real and imaginary
parts, respectively.
The one soliton solution (4) for $(q_1,q_2)$ is characterized by three arbitrary
complex parameters $\alpha_1^{(1)}$, $\alpha_1^{(2)}$, and $k_1$, in
addition to the real coupling parameters $\rho$ and $\chi$.
Also note that the value of $\chi$ is restricted by Eq. (3b) as
$|\chi|\le|\rho|$ since $\tanh{\theta}=\chi/\rho $. As in the case of mixed CNLS equations, solution (4)
can be both singular and nonsingular. The condition for
non-singular solution is given by
 $|\alpha_1^{(1)}|$ $>$ $|\alpha_1^{(2)}|$ \cite{ref20}. In this work we deal with nonsingular solutions only
as they are of specific physical interest.

\subsubsection{Analysis on bright one-soliton solution}
Typical  plot of non-singular bright one soliton
solution (4) of Eq. (2) with the condition $|\chi|\le|\rho|$ is shown in Fig. 2. From the figure we observe that 
the role of 
the linear coupling terms in (2) is to induce spatially periodic intensity switching
between the two components $q_1$ and $q_2$. The periodic oscillations Fig. 2(a) during the
intensity
switching depends particularly on the difference between the self and
cross coupling terms $(\rho$ and $\chi)$ in addition to the soliton parameters $k_1$,  
$\alpha_1^{(1)}$ and $\alpha_1^{(2)}$ . For comparison we have plotted the corresponding one-soliton solution in the absence of coupling terms in Fig. 2(b).  It is interesting to note
that this periodic intensity switching can be completely
suppressed by suitably choosing  $A_2$ or $\alpha_1^{(2)}$. 
\begin{figure}
\begin{center}
\epsfig{figure=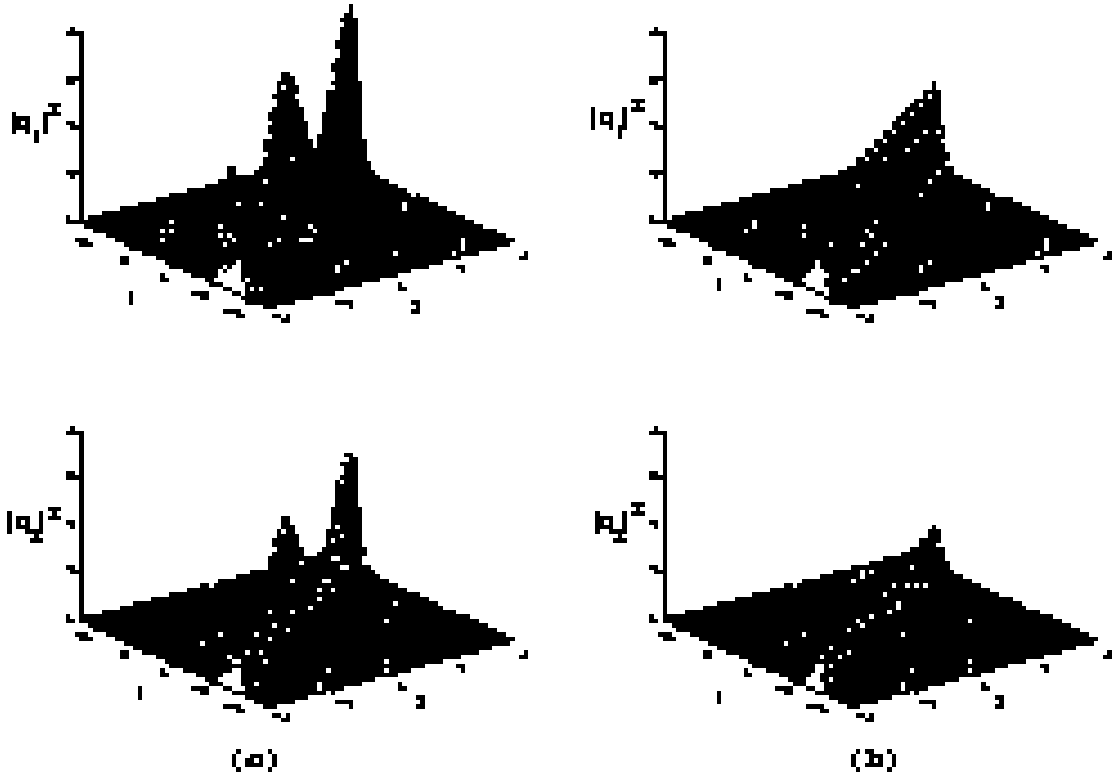, width=0.9 \columnwidth}
\caption{Intensity plots of bright one soliton solution (4):  (a) in the presence of coupling $(\rho,\chi>0)$, (b) in the absence of coupling ($\rho=\chi=0 $).}
\end{center}
\end{figure}
To see this, we compute the intensity of the soliton in the two components and write them as 
\bes
\begin{eqnarray}
\left|\frac{q_1}{P}\right|^2 &=&
k_{1R}^2\left(|A_1|^2\cosh^2\left(\frac{\theta}{2}\right)+|A_2|^2\sinh^2\left(\frac{\theta}{2}\right)\right.\nonumber\\
&&\left.
+2|A_1||A_2|\cosh\left(\frac{\theta}{2}\right)\sinh\left(\frac{\theta}{2}\right)\cos(2\Gamma
z+Q)\right),\\ 
\left|\frac{q_2}{P}\right|^2&=&
k_{1R}^2\left(|A_1|^2\sinh^2\left(\frac{\theta}{2}\right)+|A_2|^2\cosh^2\left(\frac{\theta}{2}\right)\right. \nonumber\\
&&\left.+2|A_1||A_2| \cosh\left(\frac{\theta}{2}\right) \sinh\left(\frac{\theta}{2}\right)
\cos(2\Gamma z+Q)\right),
\end{eqnarray}
where
\begin{eqnarray}
P &=& \mbox{ sech}\left(k_{1R}(t-2k_{1I}z)+\frac
{R}{2}\right),\\
Q &= & \tan^{-1}\left(
\frac{A_{1I}}{A_{1R}}\right)- \tan^{-1}\left(
\frac{A_{2I}}{A_{2R}}\right).
\end{eqnarray}
\ees
It is clear from the above expressions that the oscillatory term
$\cos(2\Gamma z+Q)$ appearing in (5a) and (5b) leads to the periodic oscillations
 during energy switching. One can also verify that the
spatial period of oscillation is $Z= \frac{\pi}{\Gamma}$. Thus for larger
$\Gamma$  the width of
spatial oscillations is smaller. Also the amplitude of oscillation ($2|A_1||A_2|\cosh\left(\frac{\theta}{2}\right)\sinh\left(\frac{\theta}{2}\right)$) increases with
decreasing $\Gamma$ due to the dependence of $\theta $ on $\chi\;\mbox{and}\; \rho $ (see Eq. (3b)) . We also note from Eqs. (5a) and (5b) that the oscillatory term (third term on the right hand side) vanishes when $|A_2| =0, \; \mbox{that is}\; \al_1^{(2)}=0$, or $|A_1| =0, \; \mbox{that is}\; \al_1^{(1)}=0$. At a first glance, it seems that the periodic energy switching scenario is similar to that of the Manakov system \cite{ma} with linear coupling terms arising in the context of twisted birefringent fibers \cite{ref22}. But the way in which the switching occurs is different due to the hyperbolic terms in Eqs. (5a) and (5b), since $0\le \sinh^2{\theta/2}< \infty , 0\le \cosh^2{\theta/2}< \infty $. To be more precise, the amplitude of periodic oscillations and periodic switching of energy between the two components can vary exponentially but restricted by the condition (3b). In this connection, we would like to add that a quite different kind of multi-scale periodic beating of intensities without switching and of different physical origin as compared with the present mixed CNLS case, has been observed during the propagation of multisoliton complexes in the integrable N-CNLS system with focusing type nonlinearities \cite{ref25}.  
\subsection{Two soliton solution}
The bright two soliton solution of Eq. (2) can be obtained by
applying the transformation (3) to the two soliton solution of the
integrable mixed CNLS equation given by Eq. (10) in Ref. \cite{ref20}. The
explicit form of the solution is 
\bes
\begin{eqnarray}
q_1 &=&\frac{1}{D}\left(
\left(\alpha_1^{(1)}e^{i\Gamma z}\cosh(\frac{\theta}{2})+
\alpha_1^{(2)}e^{-i\Gamma z}\sinh(\frac{\theta}{2})
\right)e^{\eta_1}\right. \nonumber\\ &&
\left.+\left(\alpha_2^{(1)}e^{i\Gamma z}\cosh(\frac{\theta}{2})+
\alpha_2^{(2)}e^{-i\Gamma z}\sinh(\frac{\theta}{2})
\right)e^{\eta_2}\right. \nonumber\\
&&\left.+\left(e^{\delta_{11}}\cosh(\frac{\theta}{2})e^{i\Gamma z}
+ e^{\delta_{12}}\sinh(\frac{\theta}{2})e^{-i\Gamma z}
\right)e^{\eta_1+\eta_1^*+\eta_2}\right. \nonumber\\ &&
\left.+\left(e^{\delta_{21}}\cosh(\frac{\theta}{2})e^{i\Gamma z} +
e^{\delta_{22}}\sinh(\frac{\theta}{2})e^{-i\Gamma z}
\right)e^{\eta_2+\eta_2^*+\eta_1}\right),\\
 q_2 &=&\frac{1}{D}\left(
\left(\alpha_1^{(1)}e^{i\Gamma z}\sinh(\frac{\theta}{2})+
\alpha_1^{(2)}e^{-i\Gamma z}\cosh(\frac{\theta}{2})
\right)e^{\eta_1} \right.\nonumber\\ &&
\left.+\left(\alpha_2^{(1)}e^{i\Gamma z}\sinh(\frac{\theta}{2})+
\alpha_2^{(2)}e^{-i\Gamma z}\cosh(\frac{\theta}{2})\right)
e^{\eta_2}\right. \nonumber\\ &&\left. +
\left(e^{\delta_{11}}\sinh(\frac{\theta}{2})e^{i\Gamma z} +
e^{\delta_{12}}\cosh(\frac{\theta}{2})e^{-i\Gamma z}
\right)e^{\eta_1+\eta_1^*+\eta_2}\right. \nonumber\\ && \left. +
\left(e^{\delta_{21}}\sinh(\frac{\theta}{2})e^{i\Gamma z} +
e^{\delta_{22}}\cosh(\frac{\theta}{2})e^{-i\Gamma z}
\right)e^{\eta_2+\eta_2^*+\eta_1}\right),
\end{eqnarray}
where $\alpha_l^{(j)}$'s are complex parameters and the denominator $D$ is given by
\begin{eqnarray}
D =  1+e^{\eta_1+\eta_1^*+R_1}
+e^{\eta_1+\eta_2^*+\delta_0}
 +e^{\eta_1^*+\eta_2+\delta_0^*}+e^{\eta_2+\eta_2^*+R_2}
+e^{\eta_1+\eta_1^*+\eta_2+\eta_2^*+R_3}.  \label{9b}
\end{eqnarray}
Various quantities found in Eq. (6) are defined as below following Ref. \cite{ref20}:
\begin{eqnarray}
\eta_i&=&k_i(t+ik_iz),\;\;
e^{\delta_0} = \frac{\kappa_{12}}{k_1+k_2^*},\;\;
e^{R_1} = \frac{\kappa_{11}}{k_1+k_1^*},\;\;\;\;
e^{R_2}=  \frac{\kappa_{22}}{k_2+k_2^*},\nonumber\\
e^{\delta_{1j}}&=&\frac{(k_1-k_2)(\alpha_1^{(j)}\kappa_{21}
-\alpha_2^{(j)}\kappa_{11})}{(k_1+k_1^*)(k_1^*+k_2)},\;\;
e^{\delta_{2j}}=
\frac{(k_2-k_1)(\alpha_2^{(j)}\kappa_{12}-\alpha_1^{(j)}\kappa_{22})}
{(k_2+k_2^*)(k_1+k_2^*)},\nonumber\\
e^{R_3}&=&  \frac{|k_1-k_2|^2}{(k_1+k_1^*)(k_2+k_2^*)|k_1+k_2^*|^2}
 (\kappa_{11}\kappa_{22}-\kappa_{12}\kappa_{21}), \label {rc1}
 \eear
\noindent and
\bear
\kappa_{ij}= \frac{\mu\left(\s_1\alpha_i^{(1)}\alpha_j^{(1)*}+\s_2\alpha_i^{(2)}\alpha_j^{(2)*}\right)}
{\left(k_i+k_j^*\right)},\;i,j=1,2, \label{10d}
\end{eqnarray}
\ees
where $\s_1 = 1$ and $\s_2=-1$. This solution
represents the interaction of two bright solitons in the presence of
self and cross coupling terms. Although the above solution features both
singular and nonsingular solutions in the following we consider only the
nonsingular soliton solution which results for the choice \cite{ref20}
\bes 
\begin{eqnarray}
{\kappa}_{11} \ge 0,\;\;{\kappa}_{22} \ge 0, \;\;
{\kappa}_{11}{\kappa}_{22} - |{\kappa}_{12}|^2 > 0,\\
\frac{1}{2}\sqrt{\frac{\kappa_{11}\kappa_{22}}{k_{1R}k_{2R}}}+ \frac{|k_1-k_2|}
{2|k_1+k_2^*|}\sqrt{\frac{{\kappa}_{11}{\kappa}_{22} - |{\kappa}_{12}|^2 }{{k_{1R}k_{2R}}}}
> \frac{|\kappa_{12}|}{|k_1+k_2^*|}. 
\end{eqnarray}
\ees
and analyse their collision behaviour. In a similar way the multi-soliton
solution  of Eq. (2) can
be obtained by applying the transformation to the multi-soliton solution given
in the appendix of \cite{ref20} with
$N=2$.

\section{Shape changing collision of solitons with periodic energy switching}
We have already explained the nature of type-II SCC of bright solitons in Sec. II.
In this section, we analyse the influence of linear cross coupling
terms on the above mentioned type-II SCC. We perform an asymptotic
analysis \cite{ref20} for the choice $k_{1R},  k_{2R} > 0$ and $k_{1I} >
k_{2I}$. To facilitate the understanding of the collision dynamics we consider 
the intensities of the two colliding solitons 
in the asymptotic limits at 
 $z \rightarrow -\infty $ (before collision) and $z \rightarrow \infty $ (after collision). 
In their explicit forms
the intensities of solitons as $z \rightarrow \pm \infty $ read as
\bes
\begin{eqnarray}
\left|\frac{q_j^{n\pm}}{P_n}\right|^2 &=& k_{nR}^2\left(
|A_1^{n\pm}|^2\cosh^2\left(\frac{\theta}{2}\right)
+|A_2^{n\pm}|^2\sinh^2\left(\frac{\theta}{2}\right)\right.\nonumber\\
&&\left.+2|A_1^{n\pm}||A_2^{n\pm}|\cosh\left(\frac{\theta}{2}\right)
\sinh\left(\frac{\theta}{2}\right)\cos(2\Gamma
z+Q_n)\right),
\end{eqnarray}
where
\begin{eqnarray}
P_n &= & \mbox{sech}\left(k_{nR}(t-2k_{nI}z+R_n)\right),\\
Q_n &=& \tan^{-1}\left(\frac{A_{1I}^{n\pm}}{A_{1R}^{n\pm}} \right)
-tan^{-1}\left(\frac{A_{2I}^{n\pm}}{A_{2R}^{n\pm}}
\right),\;\;n=1,2, j=1,2. 
\end{eqnarray}
\ees
Here the quantities $A_j^{n-} k_{nR}$ and $A_j^{n+} k_{nR},\; j,n=1,2,$ are the amplitudes of the ${n}$th soliton in ${j}$th component before and after interaction respectively in the absence of linear coupling ($\chi=\rho=0$), where $A_j^{n}$'s take the following forms before and after interaction \cite{ref20}:
 \begin{subequations}
\label{asy}
\bear
\left(
\begin{array}{c}
A_1^{1-}\\
A_2^{1-}
\end{array}
\right) & = & 
\left(
\begin{array}{c}
\alpha_1^{(1)} \\
\alpha_1^{(2)}
\end{array}
\right)
\frac{e^{-R_1/2}}{(k_1+k_1^*)}, \\
\left(
\begin{array}{c}
A_1^{2-}\\
A_2^{2-}
\end{array}
\right)&=&
\left(
\begin{array}{c}
 e^{\delta_{11}}\\
e^{\delta_{12}}
\end{array}
\right) \frac{e^{-(R_1+R_3)/2}}{(k_2+k_2^*)}, \\
\left(
\begin{array}{c}
A_1^{1+}\\
A_2^{1+}
\end{array}
\right)&=&
\left(
\begin{array}{c}
 e^{\delta_{21}}\\
e^{\delta_{22}}
\end{array}
\right)
\frac{e^{-(R_2+R_3)/2}}{(k_1+k_1^*)},\\
\left(
\begin{array}{c}
A_1^{2+}\\
A_2^{2+}
\end{array}
\right) &=&
\left(
\begin{array}{c}
\alpha_2^{(1)}\\
\alpha_2^{(2)}
\end{array}
\right) 
\frac{e^{-R_2/2}}{(k_2+k_2^*)}.
\eear
\ees
Various quantities occuring in Eqs. (\ref{asy}) are defined in Eq. (6).  From Eqs. (8) it can also be verified that
\begin{eqnarray}
\left|\frac{q_1^{n\pm}}{P_n}\right|^2-\left|\frac{q_2^{n\pm}}{P_n}\right|^2=
k_{nR}^2\left(|A_1^{n\pm}|^2-|A_2^{n\pm}|^2\right) = \frac{k_{nR}^2}{\mu},\quad n=1,2,
\end{eqnarray}
which is a consequence of the conservation law of
intensities in the mixed CNLS system.
\begin{figure}
\begin{center}
\epsfig{figure=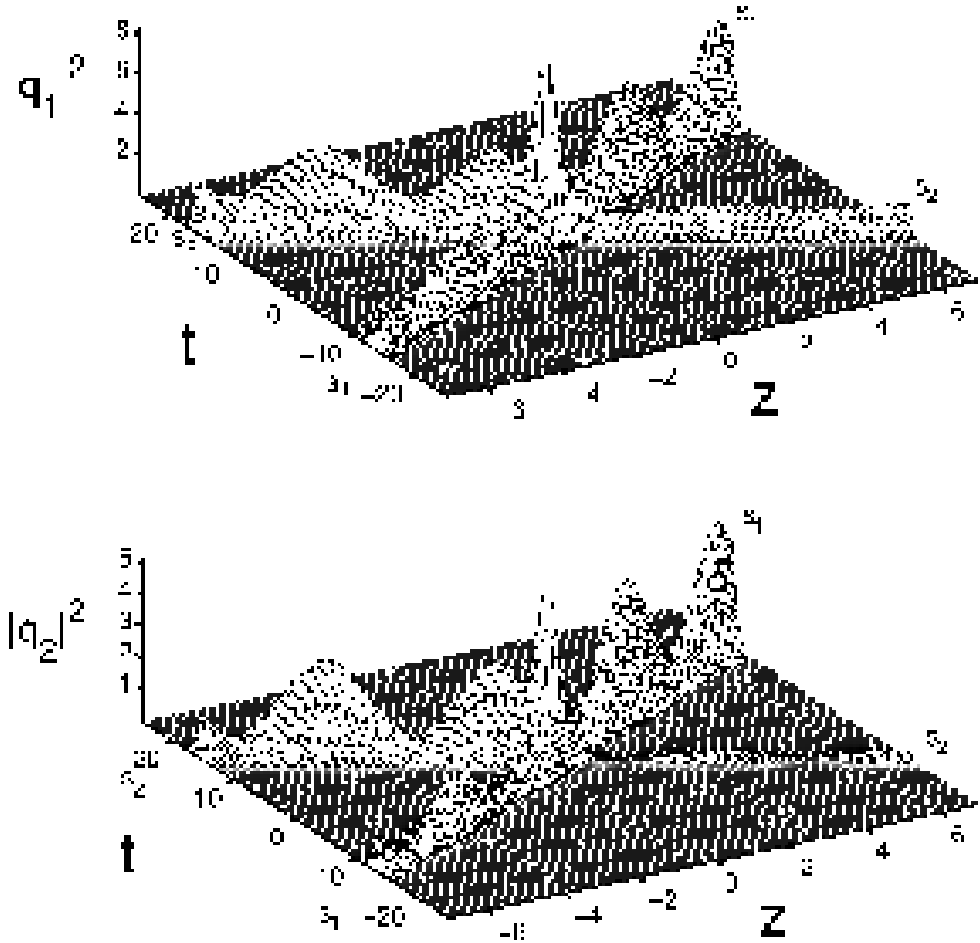, width= 0.8 \columnwidth}
\caption{Type-II shape changing collisions with periodic intensity switching.}
\end{center}
\label{a1}
\end{figure}
The role of linear coupling parameters on type-II SCC and vice-versa can
be well understood by  analysing the asymptotic
expressions (8) which clearly shows 
that these terms induce periodic switching of
intensity between the two colliding solitons in both the
components $q_1$ and $q_2$. At a first sight it seems that the periodic
intensity switching in a given soliton (say soliton $S_1$) is influenced
only by the same soliton present in the other component. But a careful
analysis shows that the presence of the second soliton (say soliton $S_2$) 
plays an indirect but predominant role in controlling the switching
process through type-II SCC and vice-versa. Various possibilities of such
collision scenario are given below:

\begin{figure}
\begin{center}
\epsfig{figure=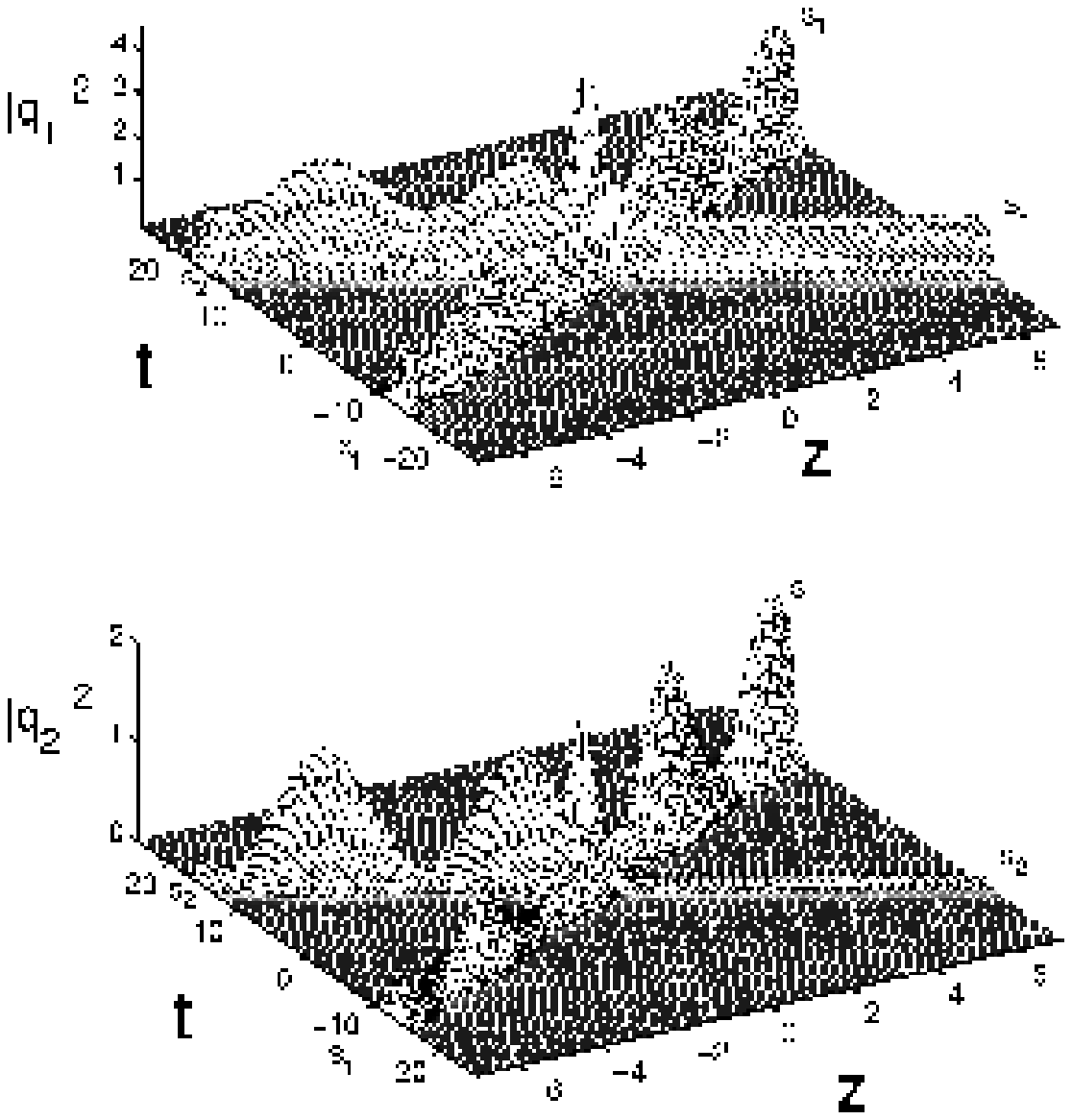, width= 0.8 \columnwidth}
\caption{Suppression of periodic oscillations in $S_2$ after interaction in a type-II SCC process.}
\end{center}
\end{figure}
\begin{figure}
\begin{center}
\epsfig{figure=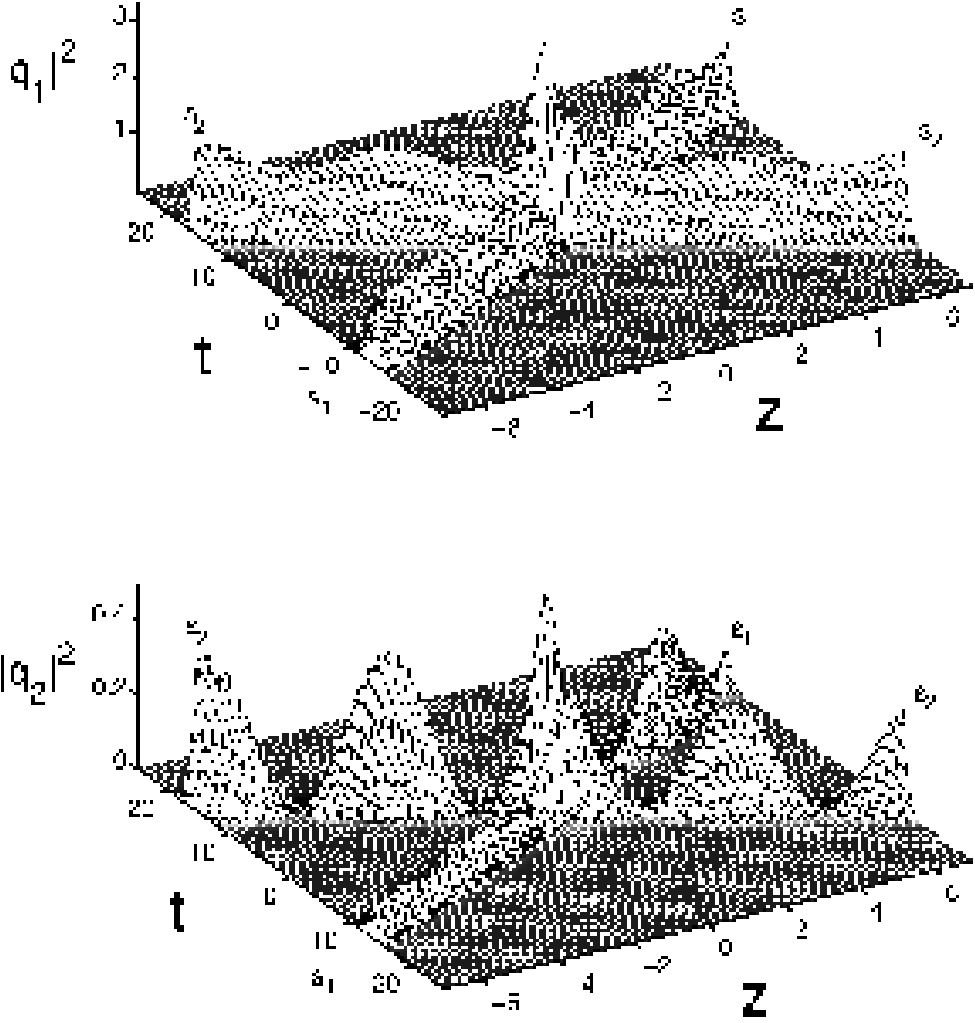, width= 0.8 \columnwidth}
\caption{Suppression of periodic oscillations in $S_1$ before interaction in a type-II SCC process.}
\end{center}
\end{figure}
\begin{figure}
\begin{center}
\epsfig{figure=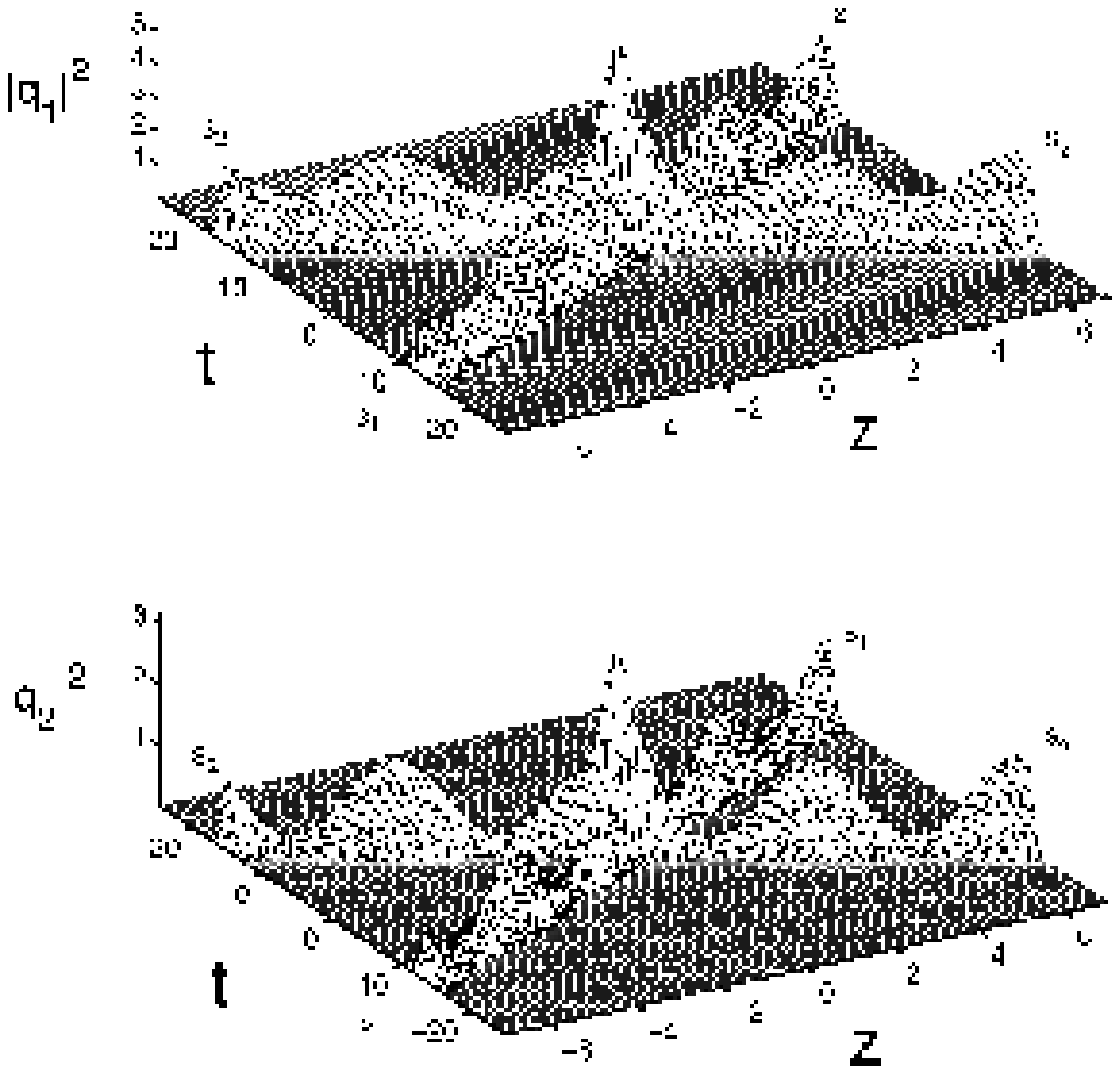, width= 0.8 \columnwidth}
\caption{Elastic collision  of bright solitons with periodic oscillations in mixed CNLS system with linear self and cross couplings. }
\end{center}
\end{figure}
\begin{enumerate}
\item
The coupling  results in periodic oscillations in the
energy switching process throughout the collision process due to
the oscillatory term $\cos(2\Gamma z+Q_n)$ appearing in Eq. (8a). As in the
case of one soliton solution here also the amplitude and width of the periodic
oscillations increase with decreasing $\Gamma$.  Thus the important feature of such collision process is that the amplitude of periodic energy switching can be large depending upon the relative signs of linear coupling terms $ \rho$ and $ \chi$.  This periodic
energy switching behaviour, in the presence of coupling, depends on the
$\alpha$ and $k$-parameters and also on the linear coupling coefficients. Thus
the oscillating energy switching process co-exists with type-II
SCC for $\frac{\alpha_1^{(1)}}{\alpha_2^{(1)}}\neq
\frac{\alpha_1^{(2)}}{\alpha_2^{(2)}}$. Such a two soliton
collision process with periodic intensity switching is shown in
Fig. 3  for  $\alpha_1^{(1)}=0.7226+1.1254i$, $\alpha_1^{(2)}=0.8484+0.2625i$,
$\alpha_2^{(1)}=0.5511+0.8584i$,   $\alpha_2^{(2)}=0.1923+0.0595i$, $\rho=1$, 
$\chi=0.5$, $k_1=1+i$, $k_2=1.1-i$. Eq. (8) also shows that the coupling
enhances the amplitude of the soliton in a given component before
and after interaction due to the contribution from the other
component  as compared with the bright soliton collision case in the absence of coupling.

\item
The distinct feature of this collision process is that the
intensity redistribution can be used to control the switching
dynamics. One interesting possibility is complete suppression of
oscillation either before or after collision in a particular soliton say
$``S_n"$ by making any one of $|A_j^{n-}|$ or $|A_j^{n+}|,\quad j,n=1,2,$ to
be zero, respectively, with commensurate changes in the other
soliton.  As the nonsingular condition (7a) of the solution rules out the possibility of making $|A_1^{n\pm}|$ to be zero , the complete suppression of periodic oscillation of intensities in both the components of soliton $S_n$ before (after) collision can be obtained by choosing $|A_2^{n-}|\; (|A_2^{n+}|)=0$.  This suppression (enhancement) of intensities of a particular soliton in a given component during the type-II SCC results in the enhancement (suppression) of amplitude of periodic oscillations  in the other colliding soliton as inferred from Eq. (8).  Fig. 4.  shows the type-II SCC scenario in which the
oscillations in the $q_1$ and $q_2$ components of $S_2$ after
interaction are completely suppressed, for the choice 
$\alpha_1^{(1)}$=0.6093+0.9489i, $\alpha_1^{(2)}= 0.4978+0.1540i$, 
$\alpha_2^{(1)}$=0.5403+0.8415i, $\alpha_2^{(2)}=0$, $\rho=1$, 
$\chi=0.5$, $k_1=1+i$, $k_2=1.1-i$. The reason for this is that in the absence
of coupling terms soliton $S_2$ undergoes type-II SCC with $S_1$ and its intensity in $q_2$ component  after interaction is  exactly zero for the given parametric
choice. Similarly Fig. 5  shows the suppression of oscillations in
$q_1$ and $q_2$ components of $S_1$ before interaction for the parametric choice $\alpha_1^{(1)}=1$, $\alpha_1^{(2)}=0$,
$\alpha_2^{(1)}=1.0201$,   $\alpha_2^{(2)}=0.2013$, $\rho=1$, 
$\chi=0.5$, $k_1=1+i$, $k_2=1.1-i$.  This kind
of switching process arises from the fact that in the absence of
coupling the intensity of $S_1$ in $q_2$ component (that is, $|A_2^{1-}|^2 k_{1R}^2$) is zero before it collides with $S_2$. 

\item
The standard elastic collision with periodic energy
switching only arises for the choice
$\frac{\alpha_1^{(1)}}{\alpha_2^{(1)}}=
\frac{\alpha_1^{(2)}}{\alpha_2^{(2)}}$. This is shown in Fig. 6 
for the parametric choice $\alpha_1^{(1)}=0.6782+1.0562i$, $\alpha_1^{(2)}=0.6782+1.0562i$, $\alpha_2^{(1)}=0.7247+0.2242i$,   $\alpha_2^{(2)}=0.7247+0.2242i$, 
$\rho=1$, $\chi=0.5$, $k_1$=$1+i$, $k_2=1.1-i$. 
\end{enumerate} 
In order to appreciate the significance of the present system, we compare the soliton collision behaviour with that of twisted birefringent fibers \cite{ref22} which involve focusing type nonlinearities. The crucial difference follows from Eq. (9), which says that the energy exchange between the two components ($q_1,q_2$) before and after collision is constant and as a result a given soliton experiences the same effect (either suppression or enhancement of intensity) in both the components during its collision with other soliton contrary to the twisted birefringent system.  Thus the amplitude of oscillation due to coupling can be simultaneously enhanced/suppressed after collision in both the components as a consequence of type II-SCC, a situation which is not possible in twisted birefringent fibers. Another important advantage is the efficiency of switching due to linear couplings. Here the coupling terms influence the energy switching exponentially due to the hyperbolic terms (see Eq. (9)). This suggests large switching of energy with small self coupling strengths, as compared with Manakov system with linear couplings, a desirable property in fiber couplers.
\section{Conclusion}
In  this paper, we have shown that the set of mixed 2-CNLS equations with linear self
and cross coupling terms can be transformed to  the standard
integrable mixed 2-CNLS equations by performing the transformation
(3a). The bright soliton solutions are obtained by applying this
transformation to the recently reported bright soliton solutions
of the mixed 2-CNLS equations \cite{ref20} without linear coupling terms. Our 
study shows that inclusion of linear self and cross coupling terms
lead to periodic energy switching among the components. We have also pointed out that due to the exponential dependence on the coupling terms, the energy switching can be large with small coupling strengths. In a two
soliton collision process such periodic energy switching coexists with the type-II shape changing collision
behaviour. However the standard elastic collision process can take place with or without
periodic energy switching for very specific parametric choices. An
important result which follows from the present study is that the shape changing
collision of type-II can be used suitably to suppress or enhance the periodic oscillations
in the energy switching process completely or partially and also simultaneously in both the components.  These results can give further
impetus in understanding the Lindner-Fedyanin system in the continuum limit, 
 and can also find potential applications in fiber couplers and in BECs.
 \section *{ACKNOWLEDGMENTS}
T. K acknowledges the support of Department of Science and Technology, Government of India under the DST Fast Track Project for young scientists.  He is also grateful to the constant support of the Principal and Management of Bishop Heber College, Tiruchirapalli.  The work of M. V and M. L are supported by a DST-IRPHA project.  M. L is also supported by a DAE-BRNS Raja Ramanna Fellowship.

\end{document}